\documentclass[fleqn,10pt]{wlscirep}
\usepackage[utf8]{inputenc}
\usepackage[T1]{fontenc}
\usepackage{amsmath}
\usepackage{amssymb}
\usepackage{bbm}
\usepackage{url}
\usepackage{bm}
\usepackage{booktabs}
\usepackage[mathscr]{euscript}
\usepackage{setspace}
\usepackage[ruled,vlined]{algorithm2e}
\usepackage{multirow}
\usepackage{graphicx}
\usepackage{subfigure}
\usepackage{xcolor}

\definecolor{singleton}{RGB}{255, 102, 0}
\definecolor{bond}{RGB}{153, 153, 51}
\definecolor{ring}{RGB}{153, 204, 204}
\usepackage{array}
\newcolumntype{L}{>{\centering\arraybackslash}m{2.8cm}}

\newcounter{firstbib}

\title{Advanced Graph and Sequence Neural Networks for Molecular Property Prediction and Drug Discovery}

\author[1,+]{Zhengyang Wang}
\author[1,+]{Meng Liu}
\author[1,+]{Youzhi Luo}
\author[1,+]{Zhao Xu}
\author[1,+]{Yaochen Xie}
\author[1,+]{Limei Wang}
\author[1,2+]{Lei Cai}
\author[3]{Qi Qi}
\author[3]{Zhuoning Yuan}
\author[3]{Tianbao Yang}
\author[1,*]{Shuiwang Ji}
\affil[1]{Texas A\&M University, Department of Computer Science and Engineering, College Station, TX 77843, USA}
\affil[2]{Currently with Microsoft, Bellevue, WA 98004, USA}
\affil[3]{University of Iowa, Department of Computer Science, Iowa City, IA 52242, USA}
\affil[*]{Correspondence should be addressed to: sji@tamu.edu}
\affil[+]{These authors contributed equally to this work.}

\begin{abstract}
Properties of molecules are indicative of their functions and thus
are useful in many applications. With the advances of deep learning
methods, computational approaches for predicting molecular
properties are gaining increasing momentum. However, there lacks
customized and advanced methods and comprehensive tools for this
task currently. Here we develop a suite of
comprehensive machine learning methods and tools spanning different
computational models, molecular representations, and loss functions for molecular
property prediction and drug discovery. Specifically, we
represent molecules as both graphs and sequences. Built on these
representations, we develop novel deep models for learning
from molecular graphs and sequences. In order to learn effectively from highly imbalanced datasets, we develop advanced loss functions that optimize areas under precision-recall curves. Altogether, our work not only
serves as a comprehensive tool, but also contributes towards
developing novel and advanced graph and sequence learning
methodologies. Results on both online and offline antibiotics
discovery and molecular property prediction tasks show that
our methods achieve consistent improvements over prior methods. In particular, our methods achieve \#1 ranking in terms of both ROC-AUC and PRC-AUC on the AI Cures Open Challenge for drug discovery related to COVID-19. Our software is released as part of the MoleculeX library under AdvProp.
\end{abstract}

\begin{document}
\flushbottom
\maketitle
\doublespacing
\section{Introduction}

Molecular property prediction is one of the key tasks in cheminformatics and has applications in many fields, including quantum mechanics, physical chemistry, biophysics and physiology~\cite{wu2018moleculenet}. With the rapid advances of machine learning methods, there is a growing trend and success to apply computational methods to predict molecular properties, which saves a huge amount of time-consuming and expensive chemical processes~\cite{schneider1998artificial,behler2007generalized,bartok2010gaussian,varnek2012machine,bartok2013representing,duvenaud2015convolutional,gilmer2017neural,schutt2017quantum,hy2018predicting,unke2019physnet,yang2019analyzing}. With emerging global challenges like the COVID-19 pandemic, it is even more urgent to develop powerful machine learning methods for molecular property prediction, thereby accelerating the speed of drug discovery~\cite{unterthiner2014deep,ma2015deep,stokes2020deep,aicures}.

In the literature, there are two common ways to represent molecules \emph{in silico}, \emph{i.e.}, molecular graphs and simplified molecular-input line-entry system (SMILES) sequences~\cite{weininger1988smiles}. Molecular graphs are intuitive to humans and informative to machine learning models. In particular, molecular graphs can keep the structure information of molecules by clearly showing the connections between atoms. On the other hand, SMILES sequences have a longer history in cheminformatics and have been widely used. SMILES is a linguistic construct to represent molecules as single-line sequences using certain simple grammar rules and a vocabulary composed of characters representing atoms and bonds. The SMILES sequences also stores essential chemical information of molecules. Various machine learning methods have been proposed on either molecular graphs~\cite{ShervashidzeSLMB11,duvenaud2015convolutional,wu2018moleculenet,kearnes2016molecular,gilmer2017neural,coley2017convolutional,schutt2017quantum,yang2019analyzing,Fey/etal/2020} or SMILES sequences~\cite{lodhi2002text,cao2012silico,bjerrum2017smiles,mayr2018large,wang2019smiles,honda2019smiles}. Setting aside the difference in data types, these machine learning methods can also be divided by the method type. That is, we can categorize them into deep learning models and non-deep learning models. Deep learning methods~\cite{lecun2015deep} have set new records in most machine learning tasks. However, non-deep learning methods still show advantages in certain cases, especially when the amount of available labeled data is small. Nevertheless, there is no clear evidence on which data type or method type leads to better performances for molecular property prediction~\cite{mayr2018large,wu2018moleculenet,yang2019analyzing}. In addition, the development of deep learning models for molecular property prediction is still in its early stage.

\begin{figure}
    \centering
    \includegraphics[width=1.0\textwidth]{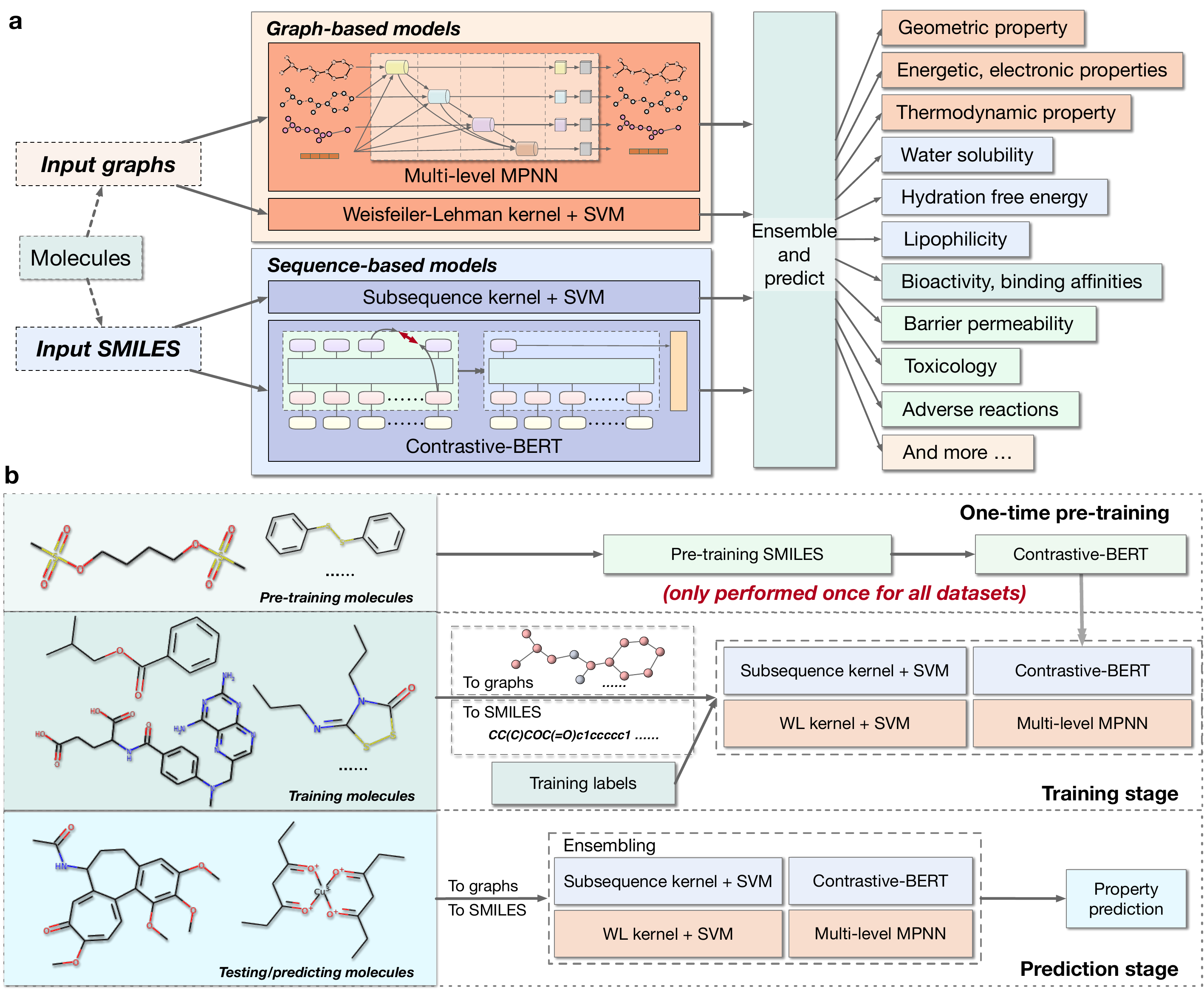}
    \caption{\textbf{AdvProp overview, training and inference.} \textbf{a}, AdvProp is composed of four modules, including two graph-based models, multi-level message passing neural network (ML-MPNN) and Weisfeiler-Lehman subtree (WL-subtree) kernel, and two SMILES sequence-based models, contrastive-BERT and subsequence kernel. We represent molecules as molecular graphs and SMILES sequences, and feed into the corresponding modules. The final prediction result of AdvProp is obtained by ensembling the predicted scores of the four modules. \textbf{b}, Given a training dataset with labeled molecules, each module is trained individually under the standard supervised setting. In particular, contrastive-BERT initializes the training parameters differently from the other three modules who apply the random initialization. The initialization of contrastive-BERT comes from an extra one-time pre-training phase on a large pre-training dataset with unlabeled molecules. In the prediction stage, AdvProp predicts the property of molecules by ensembling scores predicted by the four modules.}
    \label{fig:pipeline}
\end{figure}

In this work, we propose AdvProp, an advanced machine learning tool for molecular property prediction and drug discovery. AdvProp is unique in four aspects.
First, AdvProp consists of a suite of comprehensive machine learning methods across different data types and method types. We expect them to provide complementary information for molecular property prediction and yield better performance.
Second, we propose a new graph-based deep learning method named multi-level message passing neural network (ML-MPNN) (Section~\ref{sec:ML-MPNN}), which is included in AdvProp. ML-MPNN is able to aggregate information from a full range of levels in molecular graphs, including nodes, edges, subgraphs, and the entire graph and makes full use of richly informative molecular graphs. ML-MPNN is shown to be effective and can serve as the new state-of-the-art graph-based deep learning method for molecular property prediction.
Third, AdvProp is equipped with a novel sequence-based deep learning method, namely contrastive-BERT (Section~\ref{sec:contrastive-Bert}). Contrastive-BERT is based on the state-of-the-art BERT~\cite{devlin2018bert} and includes our proposed masked embedding recovery task, a novel self-supervised pre-training task via contrastive learning. This enables contrastive-BERT to make use of large amounts of unlabeled molecules, leading to competitive performances on downstream tasks. Fourth, AdvProp employs effective stochastic optimization methods for optimizing areas under curves (AUC), including areas under receiver operating characteristic (ROC) curves and areas under Precision-Recall curves (PRC). It is demonstrated that they can significantly improve the prediction performance in terms of ROC-AUC and PRC-AUC (Section~\ref{subsec: metrics}) on highly imbalanced datasets. By combining the advanced AUC optimization methods with our ML-MPNN models and contrastive-BERT models, we achieve the \#1 ranking on the AI Cures Open Challenge~\cite{aicures}.
With these four features, AdvProp not only serves as a comprehensive tool for molecular property prediction, but also contributes towards developing novel and advanced techniques. Concretely, AdvProp has four modules covering deep and non-deep methods based on both molecular graphs and SMILES sequences; those are, ML-MPNN, Weisfeiler-Lehman subtree (WL-subtree) kernel~\cite{ShervashidzeSLMB11}, contrastive-BERT and subsequence kernel~\cite{lodhi2002text} (Fig.~\ref{fig:pipeline}a).

In the following, we (1) demonstrate the overall effectiveness of AdvProp in the AI Cures open challenge task~\cite{aicures}, (2) show the advantages of AdvProp and each module of it in drug discovery with an off-line dataset, (3) analyze the improvements brought by AdvProp as well as our proposed ML-MPNN and contrastive-BERT on a wide range of molecular property prediction tasks from MoleculeNet benchmarks~\cite{wu2018moleculenet}. In addition, we discuss the key differences of ML-MPNN and contrastive-BERT from related works, and perform comprehensive ablation studies to support the novel designs in ML-MPNN and contrastive-BERT.

\section{Results}\label{sec:results}

\subsection{Pipeline: training and inference}

The proposed AdvProp consists of four modules; namely, multi-level message passing neural network (ML-MPNN), Weisfeiler-Lehman subtree (WL-subtree) kernel~\cite{ShervashidzeSLMB11}, contrastive-BERT and subsequence kernel~\cite{lodhi2002text} (Fig.~\ref{fig:pipeline}a). Despite the differences in data types and model architectures among these four modules, the training and inference procedures are unified (Fig.~\ref{fig:pipeline}b). First, representing molecules as molecular graphs and SMILES sequences~\cite{weininger1988smiles} is effortless. Next, in the training stage, we assume a training dataset for each molecular property prediction task, where molecules are labeled by the property of interest. Each module is then trained under the standard supervised learning setting, just with different initialization methods. Specifically, contrastive-BERT initializes training parameters from an extra one-time pre-training phase, while the other three modules apply the random initialization. The pre-training of contrastive-BERT is performed under a self-supervised contrastive learning setting on 2 million unlabeled molecules from the ZINC~\cite{irwin2012zinc} database. It is worth emphasized that this pre-training phase is conducted only once for all tasks. Finally, during inference, we use the ensemble of the predicted scores from the four modules as the final prediction result of our AdvProp. Users of AdvProp are also given the freedom of employing fewer modules.

\subsection{AI Cures open challenge task}

The COVID-19 pandemic has posed a great challenge to public health, and we first demonstrate how our AdvProp can be used to accelerate drug discovery for emerging global threats like the COVID-19.
To this end, we use our AdvProp to participate in an open challenge on drug discovery, known as the AI Cures~\cite{aicures}, which aims at discovering new antibiotics to cure the secondary lungs infections caused by COVID-19.
The entire dataset consists of 2,335 chemical molecules, which is split into a training set of 2,097 molecules and a test set of 238 molecules. The prediction target is whether a molecule has the antibacterial activity, or inhibition, to \textit{Pseudomonas aeruginosa}, which is a bacterium leading to secondary lungs infections of COVID-19 patients. The ground truth binary labels of all molecules are obtained from \textit{in-vitro} screening assay. The class labels of the training data have been made available to challenge participants, while the labels of test data have been hold out for the challenge organizers to evaluate the performance of participants. The prediction performance is evaluated by ROC-AUC and PRC-AUC (Section~\ref{subsec: metrics}) on the hold-out test set.

We use the contrastive-BERT and ML-MPNN models of AdvProp in this task. Because the data is highly imbalanced, only using the binary cross entropy loss to train the models does not result in good ROC-AUC or PRC-AUC on the test set. Inspired by the recent success of deep AUC maximization on imbalanced medical images~\cite{yuan2020robust}, we use two stochastic methods to maximize the two AUC scores. Specifically, for ROC-AUC maximization we use a stochastic optimization method for optimizing the AUC margin loss~\cite{yuan2020robust}; and for PRC-AUC maximization we use a stochastic optimization method for optimizing a surrogate loss of average precision\cite{qi2021stochastic} (Section~\ref{sec:loss}).  We follow the original papers to train deep neural networks of contrastive-BERT and ML-MPNN. In particular, we employ the two-stage framework~\cite{yuan2020robust, qi2021stochastic}. In the first stage, we learn a deep neural network by optimizing the cross entropy loss for the purpose of learning the feature network. In the second stage, we randomly re-initialize the classifier layer of the learned network in the first stage and learn all layers by optimizing the surrogate losses of ROC-AUC and PRC-AUC, respectively. We refer to the methods for deep learning by optimizing the AUC scores as deep AUC maximization method below.

Our final prediction is obtained from the ensemble of  4 ML-MPNN models and 9 contrastive-BERT models. Concretely, we use the average of  predicted inhibition probabilities from 13 models as the final predicted inhibition probability. Among all the models, 1 ML-MPNN model and 2 contrastive-BERT models are trained by deep AUC maximization using the AUC margin loss and PESG optimizer~\cite{yuan2020robust}, 1 ML-MPNN model and 3 contrastive-BERT models are trained by deep AUC maximization using the squared hinge loss for constructing the surrogate of PRC-AUC and the SOAP-Adam optimizer~\cite{qi2021stochastic}, and the remaining models are trained by the binary cross entropy loss.


\begin{table}[t]
\caption{Results on the AI Cures open challenge task in terms of PRC-AUC and ROC-AUC. Our model name is AdvProp, and our team name is DIVE@TAMU.}\label{tab:aicure}
\center
\subtable[Test PRC-AUC]{
    \begin{tabular}{|c|c|c|c|c|c|}
    \hline
    Rank & Model & Author & Submissions & Test PRC-AUC \\
    \hline
    1     & AdvProp & DIVE@TAMU & 11     & 0.729\\
    \hline
    2     & MolecularG & AIDrug@PA & 9     & 0.725\\
    \hline
    3     &   -   & AGL Team & 20    &  0.702\\
    \hline
    4     &  -    & phucdoitoan@Fujitsu & 14 & 0.694\\
    \hline
    5     & GB & BI    & 6     & 0.67\\
    \hline
    6     & Chemprop ++ & AICures@MIT & 4     & 0.662 \\
    \hline
    7     &   -   & Mingjun Liu & 3     & 0.657 \\
    \hline
    8     & Pre-trained OGB-GIN (ensemble) & Weihua Hu@Stanford & 2     & 0.651 \\
    \hline
    9     & RF + fingerprint & Cyrus Maher@Vir Bio & 1     & 0.649 \\
    \hline
    10     & Graph Self-supervised Learning & SJTU\_NRC\_Mila & 3     & 0.622 \\
    \hline
    \end{tabular}
  \label{tab:prc}}
\subtable[Test ROC-AUC]{
    \begin{tabular}{|c|c|c|c|c|c|}
    \hline
    Rank & Model & Author & Submissions & Test ROC-AUC \\
    \hline
    1     & AdvProp & DIVE@TAMU & 11     & 0.957 \\
    \hline
    2     & Chemprop ++ & AICures@MIT & 4     & 0.877 \\
    \hline
    3     &  -    & phucdoitoan@Fujitsu & 14 & 0.867\\
    \hline
    4    &   -   & Gianluca Bontempi & 7     & 0.848 \\
    \hline
    5     &   -   & Apoorv Umang & 1     & 0.84 \\
    \hline
    6    & Pre-trained OGB-GIN (ensemble) & Weihua Hu@Stanford & 2     & 0.837 \\
    \hline
    7     &   -   & Kexin Huang & 1     & 0.824 \\
    \hline
    8     & Chemprop & Rajat Gupta & 7     & 0.818 \\
    \hline
    9     & MLP & IITM  & 7     & 0.807 \\
    \hline
    10     & Graph Self-supervised Learning & SJTU\_NRC\_Mila & 3     & 0.800 \\
    \hline
    \end{tabular}
  \label{tab:auc}}
\end{table}

There are a total of 27 teams participating in this open challenge. The prediction performance of our AdvProp and those of other top teams are reported in Table~\ref{tab:aicure}. In summary, our AdvProp can outperform all other teams by a large margin. Specifically, our AdvProp achieves the highest test PRC-AUC and ROC-AUC among all teams, and our test ROC-AUC outperforms the second highest test ROC-AUC by 9\%. These results indicate that our AdvProp achieves very stable and consistent performance across two different evaluation metrics. Our promising performance and turnkey solution to this open challenge on drug discovery demonstrate that AdvProp can be used for drug screening and accelerating drug discovery.

\begin{table}[t]
  \centering
  \caption{Results on the Drug Repurposing Hub (DRH) testing dataset in terms of PRC-AUC and ROC-AUC.}
    \begin{tabular}{|c|c|c|c|c|c|c|}
    \hline
    \multirow {1}[1]{*}{Model} & Test PRC-AUC & Test ROC-AUC \\
    \hline
    Baseline  & 0.777 & 0.888 \\
    \hline
    ML-MPNN   & 0.828 & 0.900 \\
    WL-subtree kernel  & 0.857 & 0.913 \\
    Contrastive-BERT  & 0.804 & 0.877 \\
    Subsequence kernel  & 0.778 & 0.878 \\
    \hline
    AdvProp  & 0.867 & 0.919 \\
    \hline
    \end{tabular}
  \label{tab:DRHresult}
\end{table}

\subsection{Antibiotic discovery from Drug Repurposing Hub}

In addition to the AI Cures open challenge task, we apply our AdvProp on another dataset to further demonstrate its advantages on accelerating drug discovery. In particular, we focus on predicting the capability of inhibiting the growth of \textit{Escherichia coli}. It will greatly help finding new antibiotics to alleviate the bacteria’s drug-resistance problem.
The training dataset is composed of 2,335 molecules with binary labels obtained through screening. The testing dataset, provided by Stoke et al.~\cite{stokes2020deep}, contains 162 molecules form Drug Repurposing Hub (DRH), among which 53 are found inhibitory against \textit{Escherichia coli}.

The baseline model in this experiment comes from Stoke et al.~\cite{stokes2020deep} as well; that is, the directed message passing neural network (D-MPNN)~\cite{yang2019analyzing} is applied. Specifically, Stoke et al. randomly split the 2,335 training molecules into a new training dataset and a validation dataset with the ratio of 9:1 with 20 different random seeds, generating 20 different train-validation splits. Then they train D-MPNN on these 20 splits independently, yielding 20 D-MPNN models. During prediction on the testing dataset, the average of predicted scores 
over these 20 models is provided as the final prediction score. The performance is evaluated by PRC-AUC and ROC-AUC (Section~\ref{subsec: metrics}).

We apply all the four modules of AdvProp on this task. In order to show the superiority of AdvProp, we do not perform the ensemble over different train-validation splits. Instead, we randomly split the 2,335 training molecules into a new training dataset and a validation dataset with the ratio of 8.5:1.5 once, and use this train-validation split only. In addition to the ensemble of all four modules, we report the performance of each module as ablation studies, indicating that the better performance of AdvProp does not merely come from the ensemble of multiple methods. The results are provided in Table~\ref{tab:DRHresult}.

Surprisingly, the subsequence kernel, without the ensemble over different train-validation splits, can already achieve the competitive performance with the baseline by itself. The sequence-based deep learning method, contrastive-BERT, outperforms the baseline by a large margin in terms of PRC-AUC while achieving a similar ROC-AUC. The slight decrease in ROC-AUC is complemented by the graph-based methods. Both graph-based methods are very effective by themselves, especially the WL-subtree kernel. As a result, with the ensemble of four modules, AdvProp improves the performance significantly.

The results clearly show three insights regarding AdvProp. First, by comparing ML-MPNN and contrastive-BERT with the baseline, we can see that our proposed deep learning methods are effective. Second, the kernel methods are very powerful, revealing why AdvProp includes them. Third, different data types and method types provide complementary information, resulting in a further improved performance with the ensemble.

\subsection{MoleculeNet benchmarks for molecular property prediction}

With two successful applications on drug discovery, we further explore AdvProp on a wider range of molecular property prediction tasks. Particularly, we apply AdvProp on MoleculeNet benchmarks~\cite{wu2018moleculenet} for molecular property prediction.

\begin{figure}[t]
    \centering
    \includegraphics[width=1.0\textwidth]{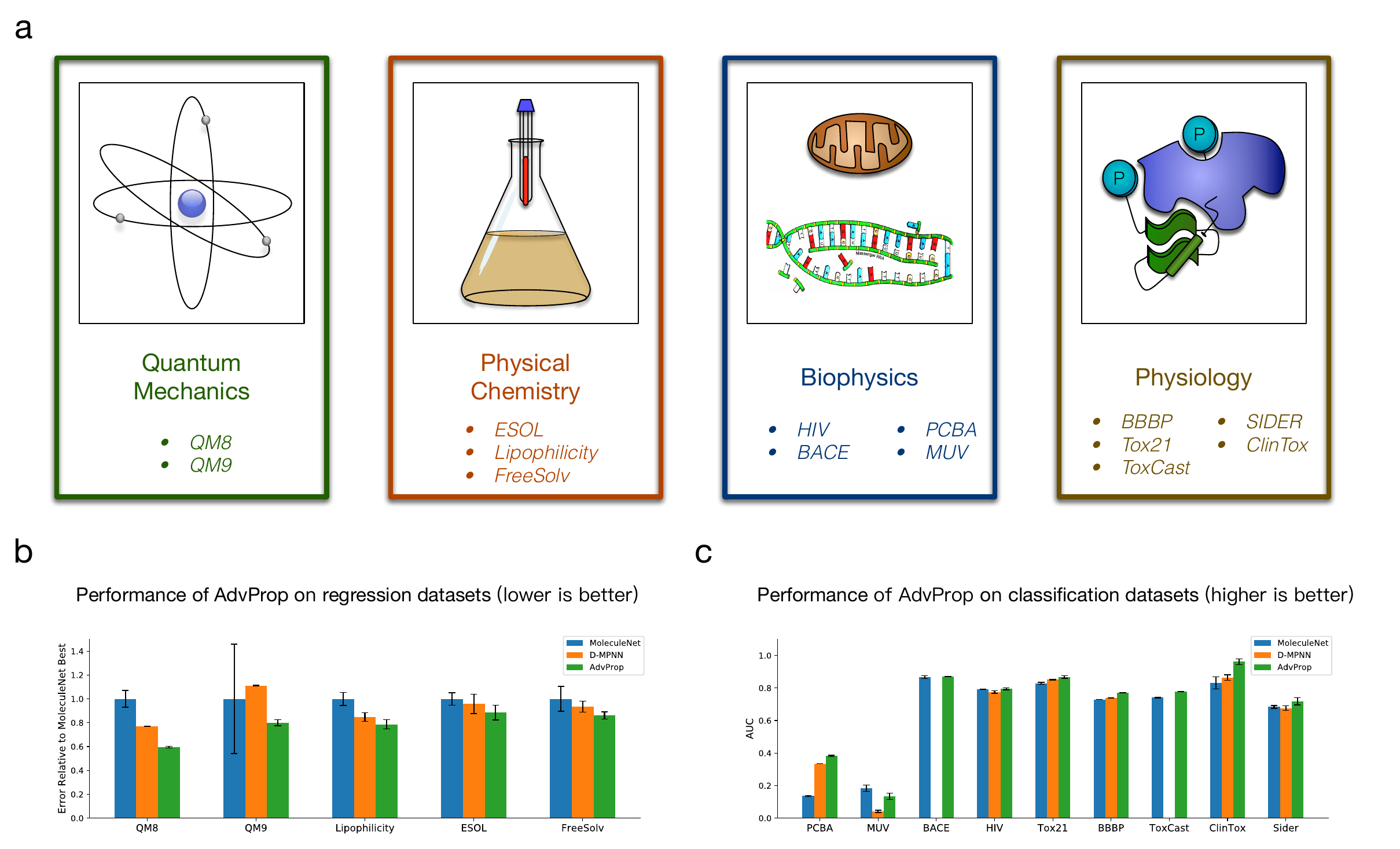}
    \caption{\textbf{AdvProp on MoleculeNet benchmarks for molecular property prediction.} \textbf{a}, We apply AdvProp on MoleculeNet benchmarks for a wide range of molecular property prediction tasks. Experiments are performed on 14 datasets, covering different molecular properties in the field of quantum mechanics, physical chemistry, biophysics and physiology. \textbf{b}, Prediction performances of AdvProp on the testing dataset of 5 regression tasks, in terms of MAE or RMSE (Section~\ref{subsec: metrics}, Supplementary Table~1). We follow Yang et al.~\cite{yang2019analyzing} to plot the error relative to the MoleculeNet baseline. The confidence intervals are marked by the standard deviations. \textbf{c}, Prediction performances of AdvProp on the testing dataset of 9 classification tasks, in terms of ROC-AUC or PRC-AUC (Section~\ref{subsec: metrics}, Supplementary Table~1). The absolute performance values are plotted. The confidence intervals are marked by the standard deviations. For b\&c, the mean and standard deviation results are computed from 3 independent runs over 3 different train/validation/test splits on each dataset. The same 3 splits are used for baselines and our AdvProp for fair comparisons. Results of MoleculeNet\cite{wu2018moleculenet} and D-MPNN\cite{yang2019analyzing} are directly copied from their papers. Note that D-MPNN is not evaluated on BACE and ToxCast datasets. All the corresponding quantitative results are provided in Supplementary Table~2. These results show significant and consistent performance gains brought by our AdvProp.}
    \label{fig:molnet_ens}
\end{figure}

We perform experiments on 14 datasets of MoleculeNet benchmarks, focusing on various molecular properties in the field of quantum mechanics, physical chemistry, biophysics and physiology (Fig.~\ref{fig:molnet_ens}a). These properties can be either binary labels, like those in drug discovery, or continuous values. Correspondingly, they are formulated as binary classification tasks or regression tasks. Typically, binary classification tasks are evaluated by PRC-AUC or ROC-AUC, while regression tasks are evaluated by MAE or RMSE (Section~\ref{subsec: metrics}). In order to serve as benchmarks and enable fair comparisons, Wu et al.~\cite{wu2018moleculenet} provide the split and evaluation methods for all the datasets. We report these key details of datasets in Supplementary Table~1. Further information can be found in Wu et al.~\cite{wu2018moleculenet}.

Two baseline models are chosen for comparisons with our AdvProp. One is a collection of previously proposed molecular featurization and learning algorithms provided together with MoleculeNet benchmarks~\cite{wu2018moleculenet}. It includes both graph-based and sequence-based methods, covering both deep and non-deep models. Without loss of clarity, we denote this baseline as MoleculeNet and always report the best performance from the collection. The other model is the directed message passing neural network (D-MPNN)~\cite{yang2019analyzing}, which incorporates latest techniques in graph neural networks and achieves the state-of-the-art performances on most datasets of MoleculeNet benchmarks. These two models form strong baselines in terms of both breadth and depth.

All the results in this section, including baselines and our AdvProp, are obtained by following the same settings provided in MoleculeNet benchmarks~\cite{wu2018moleculenet}. The performances of our AdvProp and the two baselines are compared in Fig.~\ref{fig:molnet_ens}b\&c. All the corresponding quantitative results are provided in Supplementary Table~2. Note that when showing performances on regression datasets (Fig.~\ref{fig:molnet_ens}b), we follow Yang et al.~\cite{yang2019analyzing} to plot the error relative to the MoleculeNet baseline, because the scale of performance values differ significantly across different datasets. On the other hand, when showing performances on classification datasets (Fig.~\ref{fig:molnet_ens}c), the absolute performance values in $[0,1]$ are plotted. Note that D-MPNN is not evaluated on BACE and ToxCast datasets~\cite{yang2019analyzing}. It is also worth noting that, in this set of experiments, we do not always use the ensemble of all four modules for AdvProp. Instead, we choose modules for the final ensemble based on their validation performances. Besides, the two kernel methods are not used on QM9, PCBA, MUV, ToxCast datasets, as they are too large in terms of the number of molecules or tasks (Supplementary Table~1). In these cases, the kernel methods are infeasible due to lack of memory or exceeding long computation time. Overall, our AdvProp outperforms both baselines on 13 out of 14 datasets. On the MUV dataset, AdvProp is beaten by MoleculeNet while improved D-MPNN significantly. These results demonstrate the power of AdvProp on molecular property prediction.

Additionally, in order to demonstrate the effectiveness of our proposed novel deep learning methods, ML-MPNN and contrastive-BERT, we also perform ablation studies by comparing them individually with the baselines in the following.

\begin{figure}[ht]
    \centering
    \includegraphics[width=1.0\textwidth]{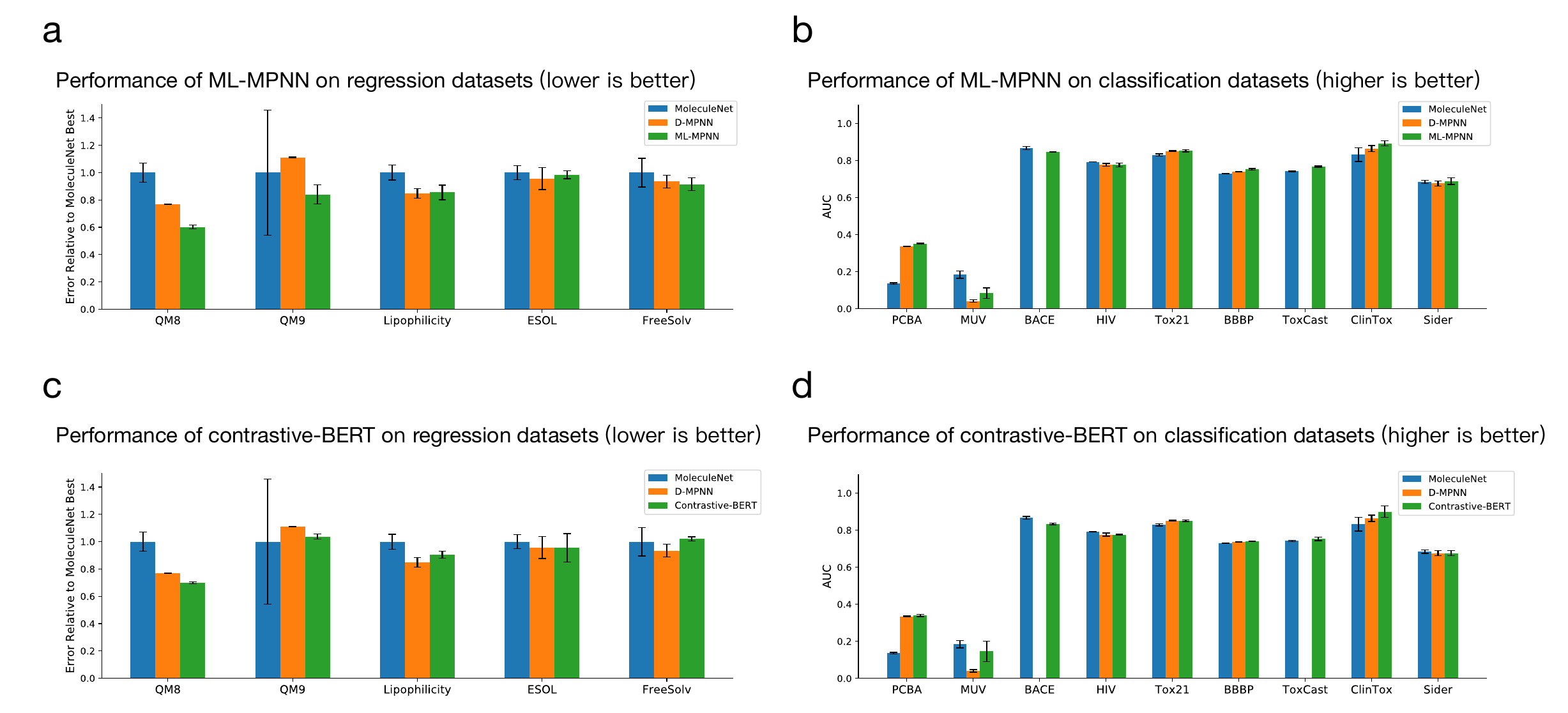}
    \caption{\textbf{ML-MPNN and contrastive-BERT on MoleculeNet benchmarks for molecular property prediction.} \textbf{a}, Prediction performances of ML-MPNN on the testing dataset of 5 regression tasks, in terms of MAE or RMSE (Section~\ref{subsec: metrics}, Supplementary Table~1). \textbf{b}, Prediction performances of ML-MPNN on the testing dataset of 9 classification tasks, in terms of ROC-AUC or PRC-AUC (Section~\ref{subsec: metrics}, Supplementary Table~1). \textbf{c}, Prediction performances of contrastive-BERT on the testing dataset of 5 regression tasks, in terms of MAE or RMSE (Section~\ref{subsec: metrics}, Supplementary Table~1). \textbf{d}, Prediction performances of contrastive-BERT on the testing dataset of 9 classification tasks, in terms of ROC-AUC or PRC-AUC (Section~\ref{subsec: metrics}, Supplementary Table~1). All the figures follow the same settings as Fig.~\ref{fig:molnet_ens}b\&c. All the corresponding quantitative results are provided in Supplementary Table~2. These results demonstrate the advantages of our proposed novel deep learning methods, ML-MPNN and contrastive-BERT.}
    \label{fig:molnet_graph_seq}
\end{figure}

\textbf{Ablation studies on ML-MPNN.} The comparisons between our proposed ML-MPNN and the two baselines, MoleculeNet and D-MPNN, are provided in Fig.~\ref{fig:molnet_graph_seq}a\&b. The plots follow the same settings as Fig.~\ref{fig:molnet_ens}b\&c. The corresponding quantitative results can be found in Supplementary Table~2.

On 9 out of 14 datasets, ML-MPNN outperforms both baselines. Concretely, ML-MPNN achieves better performances than MoleculeNet~\cite{wu2018moleculenet} on 11 out of 14 datasets and D-MPNN~\cite{yang2019analyzing} on 9 out of 12 datasets, respectively. In particular, we notice that MoleculeNet utilizes extra 3D coordinate information on QM8 and QM9 datasets, which D-MPNN and our ML-MPNN do not use. D-MPNN barely outperforms MoleculeNet on QM8 and underperforms MoleculeNet on QM9, while our ML-MPNN yields a performance boost.

The increased performances on a majority of the datasets show the effectiveness of our proposed ML-MPNN. Meanwhile, the comparisons between D-MPNN and ML-MPNN demonstrate that our proposed ML-MPNN can serve as the new state-of-the-art graph-based deep learning method for molecular property prediction.

\textbf{Ablation studies on contrastive-BERT.} The comparisons between our proposed contrastive-BERT and the two baselines are provided in Fig.~\ref{fig:molnet_graph_seq}c\&d. The corresponding quantitative results can be found in Supplementary Table~2.

On half of 14 datasets, contrastive-BERT shows performances equal to or better than both baselines. Respectively, contrastive-BERT outperforms MoleculeNet~\cite{wu2018moleculenet} on 8 out of 14 datasets and D-MPNN~\cite{yang2019analyzing} on 7 out of 12 datasets. Note that, both D-MPNN and ML-MPNN use additional global molecular features obtained with the open-source package RDKit~\cite{rdkit} (Section~\ref{subsec: config}), and MoleculeNet has access to extra 3D coordinate information on QM8 and QM9 datasets, as mentioned above. However, contrastive-BERT only uses SMILES sequences, without additional features nor extra information. Taking these into consideration, contrastive-BERT serves as a powerful sequence-based model for molecular property prediction.

\section{Discussion}

We propose AdvProp, an advanced tool composed of graph-based and sequence-based machine learning methods, for various molecular property prediction tasks including drug discovery. Specifically, AdvProp consists of four modules, including two graph-based methods, multi-level message passing neural network (ML-MPNN) and Weisfeiler-Lehman subtree (WL-subtree) kernel, and two sequence-based methods, contrastive-BERT and subsequence kernel. Among the four modules of AdvProp, ML-MPNN and contrastive-BERT are novel deep learning methods proposed in this work. We also employ advanced stochastic optimization methods for optimizing ROC-AUC and PRC-AUC measures for learning deep neural networks. These advanced deep AUC maximization methods help improve the final ROC-AUC score from 0.928 to 0.957 and improve the final PRC-AUC score from 0.677 to 0.729 on MIT AICures Challenge. In order to demonstrate the advantages of AdvProp as well as the two novel methods, we perform thorough experimental studies on two drug discovery applications and a wide range of molecular property prediction tasks.

We anticipate that our work would have potential impacts in multiple aspects. First, both our proposed ML-MPNN and contrastive-BERT achieve success in various molecular property prediction applications by themselves. They stand for advanced techniques for graph-based and sequence-based machine learning method, respectively. We expect them to not only push forward the development of new models for molecular property prediction, but also inspire the proposal of more powerful machine learning methods in general. Second, we have shown that different data types and method types both provide complementary information and boost the performances with an ensemble as AdvProp. This serves as a useful insight when applying machine learning methods for molecular property prediction in practice. In particular, our AdvProp would serve as a reliable open-source tool. Third, our work may have a larger impact beyond the academic community. For example, we achieve top performances in the AI Cures open challenge, which aims at accelerating drug discovery for emerging global threats like the COVID-19.

Regarding the two novel deep learning methods proposed in this work, ML-MPNN and contrastive-BERT, we discuss the key differences from previous studies and illustrate the superiority of our methods.
In terms of ML-MPNN, performing molecular property prediction by learning from molecular graphs has drawn a lot of attention in the community~\cite{duvenaud2015convolutional,wu2018moleculenet,kearnes2016molecular,gilmer2017neural,coley2017convolutional,schutt2017quantum,yang2019analyzing,Fey/etal/2020}. Typically, existing works learn representations for molecules through message passing. The original MPNN~\cite{gilmer2017neural} updates node representations by aggregating messages from its neighboring nodes and connected edges in each layer. Instead of updating representations for nodes, D-MPNN~\cite{yang2019analyzing} learns representations for each directed edge. The recent proposed HIMP~\cite{Fey/etal/2020} incorporates representations associated with substructures during message passing. While these existing methods are shown to be effective, they only leverage limited information contained in the molecular graphs during message passing. Intuitively, information related to the molecular properties can come from different levels, such as edges, nodes, and subgraphs. Based on these insights, our ML-MPNN is designed to be capable of aggregating information from a full range of levels in molecular graphs.

Specifically, ML-MPNN can be regarded as an extension of the unified graph neural network framework~\cite{battaglia2018relational}. The key difference lies in taking subgraph-level representations into consideration. ML-MPNN incorporates this level of information effectively, leading to a more powerful message passing framework. In order to show that the improvements of ML-MPNN do come from subgraph-level representations, we perform ablation studies to compare the performances between ML-MPNN and ML-MPNN without subgraph-level representations on MoleculeNet benchmarks. The comparison results are provided in Supplementary Table~3 and Supplementary Fig.~6. With subgraph-level representations, improvements can be observed on 12 out of 14 datasets. Another novel aspect of ML-MPNN is the normalization methods, We extend the the graph size normalization~\cite{dwivedi2020benchmarking} to SizeNorm that normalizes multi-level representations before applying BatchNorm (Section~\ref{sec:ML-MPNN}). In order to verify the effectiveness of the SizeNorm and BatchNorm, we conduct ablation studies as well and show results in Supplementary Table~4 and Supplementary Fig.~7. Compared to the model without any normalization, SizeNorm can improve the performance on 5 datasets out of 14 datasets and BatchNorm can improve the performance on 11 datasets. Applying SizeNorm and BatchNorm together can improve performance on 12 datasets.

In terms of contrastive-BERT, there exists multiple studies on learning from sequential data~\cite{hochreiter1997long,cho-etal-2014-learning,vaswani2017attention,devlin2018bert}. Recurrent neural networks (RNNs), like long short-term memory (LSTM)~\cite{hochreiter1997long} and gated recurrent unit (GRU)~\cite{cho-etal-2014-learning}, are the most popular sequence-based machine learning models before the proposal of Transformer~\cite{vaswani2017attention}. Based on Transformer, BERT~\cite{devlin2018bert} applies pre-training and achieves great success in various tasks. Transformer is theoretically more powerful than RNNs and shows improved performance in tasks like machine translation~\cite{vaswani2017attention}. However, we find it necessary to perform pre-training as in BERT to achieve good performance on molecular property prediction, due to the small sizes of available datasets. To verify this, we conduct experiments comparing performances on MoleculeNet benchmarks between GRU, Transformer, and our contrastive-BERT, and report results in Supplementary Table~5 and Supplementary Fig.~8. We can see that GRU outperforms Transformer significantly on small datasets, while contrastive-BERT achieves the best performances on 13 out of 14 datasets. Contrastive-BERT has a similar architecture as Transformer but is featured by a pre-training phase via contrastive learning~\cite{hadsell2006dimensionality,baevski2020wav2vec,he2020momentum,chen2020simple} (Section~\ref{sec:contrastive-Bert}), indicating the necessity of effective pre-training.

Contrastive-BERT is not the first study that applies BERT~\cite{devlin2018bert} on molecular property prediction. SMILES-BERT~\cite{wang2019smiles} directly employs BERT on SMILES sequences with the masked language model pre-training task from BERT. This pre-training tasks is a classification task, where the model is asked to predict what the masked characters are. On the contrary, our contrastive-BERT proposes a novel pre-training task via contrastive learning, named masked embedding recovery task, where the model is asked to predict the embeddings of the masked characters. This contrastive learning task is substantially harder than the original classification task and is supposed to encourage learning better representations of SMILES sequences for down-streaming tasks. In order to demonstrate the advantages of our proposed contrastive-BERT and the masked embedding recovery task, we perform ablation studies comparing our contrastive-BERT with SMILES-BERT model. Results are provided in Supplementary Table~6 and Supplementary Fig.~9. Our contrastive-BERT model outperforms the SMILES-BERT model on 11 out of 14 datasets.


One limitation of our AdvProp is that it does not consider the 3D coordinate information of molecules when available. While we have shown that AdvProp is able to outperform models with 3D coordinate information into consideration, we expect incorporating such information into AdvProp may further improve the performances. Intuitively, molecules are 3D structures. Location information of atoms can be important for predicting molecular properties. Although there are several attempts~\cite{schutt2017schnet,unke2019physnet,klicpera2019directional} to consider such 3D coordinate information, how to utilize such information effectively is still an open challenge in the community. We leave incorporating 3D coordinate information into AdvProp as future work.

\section*{Acknowledgements}

We thank Sheng Wang, Zhenqin Wu and Kevin Yang for sharing their data and answering our questions. This work was supported in part by National Science Foundation grants DBI-1922969, IIS-1908198, and IIS-1955189. Q.Q. Z.Y. and T.Y. are partially supported by NSF \#1933212, NSF CAREER Award \#1844403.

\section*{Author contributions}

S.J. conceived and initiated the research. Z.W., M.L., Y.L., Z.X., Y.X., L.W., L.C. and S.J. designed the methods.
M.L., Y.L., Z.X., Y.X. and L.W. implemented the training and validation methods and designed and developed the software package. Q.Q., Z.Y. and T.Y contributed to the AI Cures Open Challenge task. Z.Y. contributed to the code development of ROC-AUC optimization for deep learning and Q.Q. contributed to the code development of PRC-AUC optimization for deep learning, both under guidance of T.Y.. Z.W., L.C. and S.J. supervised the project.
Z.W., M.L., Y.L., Z.X., Y.X., L.W. and S.J. wrote the manuscript. T.Y. provided suggestions and comments on the manuscript.

\section*{Competing Interests}

The authors declare no competing interests.

\clearpage

\section{Methods}

\subsection{Notations and problem definitions}\label{sec:notations}


Our proposed AdvProp aims at making accurate predictions on certain properties of a molecule through graph-based and sequence-based machine learning models. Correspondingly, we represent a molecule as the molecular graph and the simplified molecular-input line-entry system (SMILES) sequence~\cite{weininger1988smiles}. Here, we give basic notations and define the problems.

As a molecule is composed of chemical atoms and bonds between them, it is natural to represent it as a molecular graph, where nodes and edges correspond to atoms and bonds, respectively. In particular, a bond could be represented as a single undirected edge or two directed edges with opposite directions. Formally, a molecular graph is denoted as $G = (V, E) \in \mathcal{G}$, where $V$ is the set of nodes and $E \subseteq V \times V$ is the set of edges. $|V|$ and $|E|$ are the numbers of nodes and edges, respectively. Moreover, we consider attributed molecular graphs. Specifically, there are feature vectors associated with each node $v \in V$ and edge $e \in E$. These feature vectors are represented as $x(v) \in \mathbb{R}^p$ and $w(e) \in \mathbb{R}^q$, given by the pre-defined node attribute mapping $x: V \mapsto \mathbb{R}^p$ and edge attribute mapping $w: E \mapsto \mathbb{R}^q$, respectively.

On the other hand, a molecule can be represented as a SMILES sequence as well. Basically, a SMILES sequence is simply a string, where each character denotes a chemical atom or an indicator of structures like a bond, a ring, etc. Formally, we denote a SMILES sequence as $S \in \mathcal{S}$ and its length as $|S|$. And $S[i]$, $i=1,\ldots,|S|$ refers to the $i$-th character of $S$. All the characters come from a vocabulary $\Sigma$.


A property of a molecule can be represented by a scalar number $y$. A continuous $y \in \mathbb{R}$ can denote the water solubility of the molecule. A binary $y \in \{-1, 1\}$ can indicate whether a molecule is antibacterial or not. Depending on $y$, the problem we focus on can be formulated as either a regression ($y \in \mathbb{R}$) or a binary classification ($y \in \{-1, 1\}$) task. Specifically, given a molecule represented by the molecular graph $G$ or the SMILES sequence $S$ as the input, a machine learning model is asked to predict $y$.

In the following, we describe our proposed AdvProp ensembling both graph-based and sequence-based machine learning methods.

\subsection{Graph-based machine learning methods in AdvProp}

With the molecular graph $G$ as the input, AdvProp consists of two graph-based machine learning models; those are, our novel multi-level message passing neural network (ML-MPNN) and the Weisfeiler-Lehman subtree kernel~(WL-subtree)~\cite{ShervashidzeSLMB11}.

\subsubsection{Multi-level message passing neural network}\label{sec:ML-MPNN}

Our proposed multi-level message passing neural network (ML-MPNN) is featured by the ability of aggregating information from a full range of levels in molecular graphs, including nodes, edges, subgraphs, and the entire graph. As a result, multi-level representations are initialized and updated through our ML-MPNN.

\textbf{Multi-level Representations.} We first introduce the multi-level representations and their initialization. In particular, we consider node-level representations, edge-level representations, subgraph-level representations, and a graph-level representation.

As discussed in Section~\ref{sec:notations}, the molecular graph $G = (V, E)$ is attributed, with feature vectors $x(v) \in \mathbb{R}^p$ and $w(e) \in \mathbb{R}^q$ associated with each node $v \in V$ and edge $e \in E$. The node-level representations and edge-level representations in our ML-MPNN are then initialized as $x(v) \in \mathbb{R}^p$ and $w(e) \in \mathbb{R}^q$, respectively. Concretely, we follow Yang et al.~\cite{yang2019analyzing} to define $x(\cdot)$ and $w(\cdot)$. The node attribute mapping $x(\cdot)$ encodes important attributes of a chemical atom, such as the type of atom, number of related bonds, formal charge, and atomic mass. And the edge attribute mapping $w(\cdot)$ encodes useful properties of a bond, including the bond type, the stereo, whether the bond is part of a ring, and whether the bond is conjugated. In the following, we directly use $x(v) \in \mathbb{R}^p$ and $w(e) \in \mathbb{R}^q$ to denote the node-level representations and edge-level representations for simplicity.

ML-MPNN utilizes the junction tree to obtain subgraphs and initialize the subgraph-level representations. The junction tree represents a molecular graph as a tree, where each node in the tree corresponds to a subgraph in the original molecular graph. Specifically, we follow the tree decomposition algorithm~\cite{rarey1998feature, jin2018junction, Fey/etal/2020} to build the junction tree. Nodes in the resulted junction tree represent rings, bonds, bridged compounds, or singletons in the original molecular graph, as illustrated in Supplementary Fig.~1. Afterwards, in order to initialize subgraph-level representations for the original molecular graph, we simply generate node-level representations for the junction tree. Here, for each node in the junction tree, we encode the concrete subgraph type as its node-level representation. Formally, given a molecular graph $G$, we denote the node set of the junction tree as $C$, which is also the set of subgraphs. The subgraph-level representations are denoted by $r(c) \in \mathbb{R}^f$ for each subgraph $c \in C$.


The graph-level representation in our ML-MPNN is empirically initialized as the average of edge-level representations in the corresponding graph. After ML-MPNN, the final graph-level representation obtained is used for predicting the molecular property. Formally, given a molecular graph $G$, we denote the graph-level representation as $z(G) \in \mathbb{R}^g$.

As a summary, given a molecular graph $G = (V, E)$ and the set of corresponding subgraphs $C$, the multi-level representations involved in our ML-MPNN are node-level representations $x(v) \in \mathbb{R}^p$ for each node $v \in V$, edge-level representations $w(e) \in \mathbb{R}^q$ for each edge $e \in E$, subgraph-level representations $r(c) \in \mathbb{R}^f$ for each subgraph $c \in C$, and a graph-level representation $z(G) \in \mathbb{R}^g$. Note that ML-MPNN handles molecular graphs with directed edges. For each edge $e \in E$, we use $s_e \in V$ and $t_e \in V$ to denote the source node and target node. In addition, without loss of generality, we consider the nodes and subgraphs are indexed from $1$ to $|V|$ and from $1$ to $|C|$, respectively. Then the assignment of nodes to subgraphs can be denoted by an assignment matrix $R \in \{0,1\}^{|V| \times |C|}$, where $R_{[v,c]}=1$ if node $v$ belongs to subgraph $c$, and $R_{[v,c]}=0$ otherwise.

\textbf{ML-MPNN.} Next, we introduce how multi-level representations are updated through our ML-MPNN, where message passing is performed hierarchically. ML-MPNN can be viewed as an extension of the general graph neural network framework~\cite{battaglia2018relational}. ML-MPNN stacks several ML-MPNN layers followed by a final output layer for classification or regression. Each ML-MPNN layer updates the representations for each level in five steps, as illustrated in Supplementary Fig.~2.

\textit{Step 1: Update the edge-level representations.} We update the edge-level representations by aggregating information from node-level representations of the source node and target node as well as the graph-level representation. Specifically, we concatenate these representations with the previous edge-level representation as the inputs to a multilayer perceptron (MLP) to update the edge-level representation. Formally, we have
\begin{equation}
    w'(e) = \text{MLP} \left(\left[w(e), x(s_e), x(t_e), z(G)\right]\right), \quad \forall e \in E,
\end{equation}
where $[\cdot]$ denote concatenation and $w'(e)$ represents the updated edge-level representation. Note that, as the graph-level representation contains the global information of the molecular graph, it is supposed to guide the updating process and thus used in each updating step below.

\textit{Step 2: Update the node-level representations.} For each node, we use updated edge-level representations of incoming edges and node-level representations of corresponding source nodes to compute messages it receives. The node-level representations is then updated with the messages and the graph-level representation. The updating process is formulated as
\begin{align}
    &M^{e2n}(v) = \text{MLP}\left(\sum_{\{e:\ t_e=v\}} \left[w'(e), x(s_e)\right]\right), \quad \forall v \in V, \\
    &x'(v) = \text{MLP}\left(\left[x(v), M^{e2n}(v), z(G)\right]\right), \quad \forall v \in V.
\end{align}
Here, $x'(v)$ is the updated node-level representation of node $v$. And $M^{e2n}(\cdot)$ denotes the messages passed from edge-level representations to node-level representations.

\textit{Step 3: Update subgraph-level representations.} The messages for updating the subgraph-level representations are from nodes that are assigned to the subgraph as well as neighboring subgraphs, \emph{i.e.}, neighboring nodes in the junction tree. The concrete updating procedure is given by
\begin{align}
    &M^{n2s}(c) = \text{MLP}\left(\sum_{\{v:\ R_{[v,c]}=1\}} x'(v)\right), \quad \forall c \in C, \\
    & r'(c) = \text{MLP}\left(\left[r(c), M^{n2s}(c), \text{MLP}\left(\sum_{u \in \mathcal{N}(c)} r(u)\right),  z(G)\right]\right), \quad \forall c \in C,
\end{align}
where $\mathcal{N}(c)$ denotes the set of neighboring subgraphs of subgraph $c$ and $r'(c)$ is the updated subgraph-level representation of subgraph $c$. And $M^{n2s}(\cdot)$ represents the messages passed from node-level representations to subgraph-level representations.

\textit{Step 4: Update the graph-level representation.} Finally, the graph-level representation gets updated after receiving messages from updated representations of each level. We have
\begin{align}
    & M^{e2g}(G) = \frac{1}{|E|} \sum_{e \in E} w'(e), \\
    & M^{n2g}(G) = \frac{1}{|V|} \sum_{v \in V} x'(v), \\
    & M^{s2g}(G) = \frac{1}{|C|} \sum_{c \in C} r'(c), \\
    & z'(G) = \text{MLP}\left(\left[z(G), M^{e2g}(G), M^{n2g}(G), M^{s2g}(G)\right]\right),
\end{align}
where $z'(G)$ is the updated graph-level representation of graph $G$. And $M^{e2g}(\cdot)$, $M^{n2g}(\cdot)$, and $M^{s2g}(\cdot)$ denote the messages passed from edge-level representations, node-level representations, and subgraph-level representations, respectively.

\textit{Step 5: Normalization for multi-level representations.} Like modern deep learning models, our ML-MPNN uses batch training. Adding BatchNorm~\cite{ioffe2015batch} can stabilize the training process and usually achieve better performance. However, due to the various sizes of graphs, batching graphs will lead to representations at different scales, making it difficult to learn the optimal statistics for BatchNorm~\cite{dwivedi2020benchmarking}. As a result, certain normalization methods on sizes must be applied. In ML-MPNN, graphs of various sizes refer to graphs with a variable number of edges, nodes, and subgraphs. Correspondingly, we apply the EdgeSizeNorm, NodeSizeNorm, and SubgraphSizeNorm for edge-level representations, node-level representations, and subgraph-level representations, respectively, given by
\begin{align}
    & \bar{w}'(e) = \text{BatchNorm}\left(\text{EdgeSizeNorm}\left(w'(e)\right)\right), \quad \text{EdgeSizeNorm}\left(w'(e)\right) = \frac{w'(e)}{\sqrt{|E|}}, \quad \forall e \in E, \\
    & \bar{x}'(v) = \text{BatchNorm}\left(\text{NodeSizeNorm}\left(x'(e)\right)\right), \quad \text{NodeSizeNorm}\left(x'(v)\right) = \frac{x'(e)}{\sqrt{|V|}}, \quad \forall v \in V, \\
    & \bar{r}'(c) = \text{BatchNorm}\left(\text{SubgraphSizeNorm}\left(r'(c)\right)\right), \quad \text{SubgraphSizeNorm}\left(r'(c)\right) = \frac{r'(c)}{\sqrt{|C|}}, \quad \forall c \in C, \\
    & \bar{z}'(G) = \text{BatchNorm}\left(z'(G)\right).
\end{align}

\subsubsection{Weisfeiler-Lehman subtree kernel}\label{sec:graph_kernel}

AdvProp applies the Weisfeiler-Lehman subtree kernel~(WL-subtree)~\cite{ShervashidzeSLMB11} as a traditional graph-based machine learning component. WL-subtree is able to capture the similarity among molecular graphs regarding the isomorphism. In WL-subtree, we consider molecular graphs with undirected edges. In addition, the node attribute mapping $x: V \mapsto \mathbb{R}^p$ becomes a node label mapping $\ell: V \mapsto \mathbb{Z^+}$ that assigns to each node $v$ the atomic number of the corresponding atom. And there is no edge attribute.

In general, given any two graphs $G_1, G_2\in \mathcal G$, a graph kernel computes the similarity between them as $k_g(G_1, G_2)$, where $k_g:\mathcal G^2\to \mathbb{R}$ is a kernel function. Suppose we have a training dataset $\{(G_i, y_i)\}_{i=1}^N$ with $N$ samples. We train a kernel Support Vector Machine~(SVM)~\cite{cortes1995support,chang2011libsvm} by optimizing
\begin{equation}\label{kernel_svm}
    \min_{\{\alpha_i\}_{i=1}^n}C\left[\sum_{i=1}^N L\left(\sum_{j=1}^N \alpha_j y_j K_{ji}, y_i\right)\right]+\frac{1}{2}\sum_{i=1}^N\sum_{j=1}^N\alpha_i\alpha_j K_{ji},
\end{equation}
where $K_{ij}=k_g(G_i, G_j)$, $C$ is a hyper-parameter, and $\{\alpha_i\}_{i=1}^N$ are training parameters. For a regression task, $L$ is the $\epsilon$-insensitive loss~\cite{Vapnik1995nature,drucker1997support}. After training, the prediction result of a graph $G$ is given by $\sum_{i=1}^N\alpha_i y_i k_g(G_i,G)$. For a binary classification task, $L$ is the hinge loss~\cite{cortes1995support} and the prediction result is given by $\text{sign}\left(\sum_{i=1}^N\alpha_i y_i k_g(G_i,G)\right)$.

In particular, WL-subtree defines a kernel function $k_g$ that computes the similarity based on common subtrees in two graphs. Specifically, let $\{s_1,s_2,\cdots,s_M\}$ be the set of all possible subtrees within a certain depth. For a graph $G$, we compute
\begin{equation}
\phi(G)=\left[n(s_1, G), n(s_2, G), \cdots, n(s_M, G)\right]^T,
\end{equation}
where $n(s_i, G)$, $i=1,2,\ldots, M$ is the number of the subtree $s_i$ in $G$. With $\phi(\cdot)$, the kernel function between $G_1$ and $G_2$ is given by $k_g(G_1,G_2)=\phi(G_1)^T\phi(G_2)$.

The set $\{s_1,s_2,\cdots,s_M\}$ is obtained by an iterative process of relabeling each node. Concretely, for the $t$-th iteration, we denote the label of each node $v$ in a graph $G$ as $\ell^{(t)}(v)$. At iteration 0, all the nodes in $G$ are labelled by the node label mapping $\ell^{(0)}(\cdot)=\ell(\cdot)$. At iteration $t$ ($t\geq 1)$, we update labels in two steps. First, for each node $v$, we build a multiset $S(v) = \{\ell^{(t-1)}(u):u\in\mathcal{N}(v)\}$, where $\mathcal{N}(v)$ is the set of neighboring nodes of $v$ in $G$. Basically, $S(v)$ collects the current labels of neighboring nodes of $v$. Note that a multiset is allowed to include repeated elements. Meanwhile, labels in $S(v)$ are sorted. Second, we hash $\{\ell^{(t-1)}(v), S(v)\}$ to a new label $s_v$. Then we have $\ell^{(t)}(v)=s_v$.

This iterative process is performed on all graphs in the training dataset. Note that, for any pair of different nodes $v_1$ and $v_2$, no matter whether they are in the same graph or not, $\ell^{(t)}(v_1)=\ell^{(t)}(v_2)$ if and only if the two $t$-hop subtrees rooted at $v_1$ and $v_2$ are identical. As a result, each label denotes a unique subtree. After T iterations, we collect all the labels that ever appear in any graph during the $T$ iterations and put them into $\{s_1,s_2,\cdots,s_M\}$, a set of all possible subtrees within $T$ hops. Finally, to obtain $n(s_i, G)$, $i=1,2,\ldots, M$ for any graph $G$, we simply perform the $T$ iterations on $G$ and count the times of $s_i$ appearing.

\subsection{Sequence-based machine learning methods in AdvProp}

With the SMILES sequence $S$ as the input, AdvProp consists of two sequence-based machine learning models; those are, our proposed contrastive-BERT and the subsequence kernel~\cite{lodhi2002text}.

\subsubsection{Contrastive-BERT}\label{sec:contrastive-Bert}

Our contrastive-BERT is characterized by a novel self-supervised pre-training task via contrastive learning~\cite{hadsell2006dimensionality,baevski2020wav2vec,he2020momentum,chen2020simple}, namely masked embedding recovery task. It results in better performances on downstream applications after fine-tuning. Note that the pre-training phase is performed only once on a large unlabeled dataset and we have provided the pre-trained model. The fine-tuning phase is the same as end-to-end supervised training on labeled datasets, except for using the pre-trained model to initialize training parameters instead of random initialization. In the following, we first describe the network architecture of our contrastive-BERT and then focus on the proposed masked embedding recovery task.

\textbf{Network architecture.} Similar to the original BERT~\cite{devlin2018bert}, the network architecture of our contrastive-BERT mainly follows Transformer~\cite{vaswani2017attention}. Given a SMILES sequence $S$ with the vocabulary $\Sigma$, a trainable embedding layer first is applied to transform each character $S[i]$, $i=1,2,\ldots,|S|$ into an embedding vector $h_i \in \mathbb{R}^d$. The embedding layer actually maintains an embedding matrix of dimension $|\Sigma| \times d$, where each row corresponds to the embedding vector of a character in $\Sigma$. The embedding vector of a character is supposed to capture chemical information about the corresponding atom or structure. In addition, since the order matters in SMILES sequences, the position embedding from Vaswani et al.~\cite{vaswani2017attention} is added to $h_i$. Afterwards, $h_i$ serves as the inputs to Transformer and gets updated.

Transformer is composed of a stack of Transformer layers. A Transformer layer consists of a self-attention mechanism followed by a multilayer perceptron (MLP). Specifically, with $h_i$, $i=1,2,\ldots,|S|$ as inputs, the self-attention mechanism allows direct interactions among them and updates each $h_i$ by aggregating information according to the interaction results. Formally, by writing $h_i$, $i=1,2,\ldots,|S|$ into a matrix $H=[h_1, h_2, \ldots, h_{|S|}]^T \in \mathbb{R}^{|S| \times d}$, we have
\begin{align}
& H' = \text{Softmax}\left(\frac{(HW^Q) (HW^K)^T}{\sqrt{d}}\right) HW^V,
\end{align}
where $H'=[h'_1, h'_2, \ldots, h'_{|S|}]^T \in \mathbb{R}^{|S| \times d}$ represents the updated embeddings and $W^Q, W^K, W^V \in \mathbb{R}^{d \times d}$ are training parameters. In particular, the multi-head technique~\cite{vaswani2017attention} is also used. The self-attention mechanism is similar to a global message passing in the sense that each character can receive information from any other character. The following MLP has two layers with the Gaussian error linear unit (GELU)~\cite{hendrycks2016gaussian} as the activation function. In order to facilitate the training, residual connections~\cite{he2016deep} and the layer normalization \cite{ba2016layer} are applied on both the self-attention mechanism and the MLP.

Consequently, the outputs of our contrastive-BERT are updated embedding vectors for each character of a SMILES sequence $S$, denoted as $o_i$, $i=1,2,\ldots,|S|$. In practice, a special <TASK> token is inserted to $S$ as the first character $S[1]$. The output embedding vector of <TASK> will be used for prediction, as explained below.

\textbf{Masked embedding recovery task.} As described above, the training of the contrastive-BERT consists of a self-supervised pre-training phase followed by a supervised fine-tuning phase, as shown in Supplementary Fig.~4. Our contrastive-BERT mainly contributes to the self-supervised pre-training phase by proposing the masked embedding recovery task, a novel self-supervised task via contrastive learning.

Basically, the masked embedding recovery task asks the model to predict the embeddings of masked characters based on embeddings of unmasked characters. The masking strategy in our contrastive-BERT follows the one in the original BERT~\cite{devlin2018bert}, in order to fairly demonstrate the advantages of the new objective. Concretely, in a SMILES sequence, 15\% of the characters are randomly masked. For each character that is selected to be masked, it has 80\% probability to be replaced by a special <MASK> token, 10\% probability to be replaced by another random character, and 10\% probability to remain unchanged. For each SMILES sequence, at least one character is masked.

As a result, if a character $S[i]$ is masked, the corresponding input embedding $h_i$ is replaced and will not be fed into the network. The masked embedding recovery task is to predict the output embedding $o_i$ without $h_i$. However, there is no ground truth for $o_i$, which means a common supervised objective cannot be used. In order to perform the pre-training, our contrastive-BERT designs a self-supervised objective via contrastive learning. In particular, for a masked character $S[i]$, we force the predicted $o_i$ to be similar to $h_i$, as they correspond to the same character. In contrast, $o_i$ needs to be less similar to $h_j, \forall j \neq i$. Note that, while $h_i$ is replaced from the inputs to the network, it can still be retrieved from the embedding matrix. Given a function to measure the similarity, the pre-training can be achieved using a contrastive loss, as illustrated in Supplementary Fig.~5.

After the pre-training phase with the masked embedding recovery task, our contrastive-BERT is fine-tuned on labeled datasets for downstream applications. Specifically, a predictor is added on top of the embedding vector of <TASK> to predict the molecular property. The fine-tuning is an end-to-end training process with a supervised loss. In both training phases, we use a data augmentation technique based on a characteristic of SMILES sequences. That is, one molecule may be represented by multiple different SMILES sequences, although each SMILES sequence only corresponds to a certain molecule. For example, \textit{CCN}, \textit{NCC}, and \textit{C(N)C} are all valid SMILE sequences of the organic compound ethylamine with the formula $CH_3CH_2NH_2$ (Supplementary Fig.~3). Each time we perform training on a molecule, we randomly choose a valid SMILES sequence as inputs to our contrastive-BERT. Such data augmentation helps the model learn embeddings that capture intrinsic chemical information irrelevant to the representation formats. During testing, we use the canonicalization algorithm~\cite{neglur2005assigning} to generate a unique canonical SMILES sequence for each molecule.

\subsubsection{Subsequence kernel}\label{sec:seq-kernel}

As a traditional sequence-based machine learning component, our AdvProp uses the subsequence kernel~\cite{lodhi2002text}. As indicated by its name, the subsequence kernel computes the similarity between SMILES sequences based on their common subsequences.
Formally, for a SMILES sequence $S\in\mathcal{S}$, a sequence $U$ is called a subsequence of $S$ if there exists a list of indices $\bm{i}=\{i_1,i_2,\cdots,i_{|U|}: 1\le i_1<i_2<\cdots<i_{|U|}\le |S|\}$ such that $U[j]=S[i_j]$, $j=1,\cdots,|U|$, \emph{i.e.}, $U=S[\bm i]$. Note that the subsequence $U$ is not necessarily composed of consecutive characters from $S$.

Similar to the graph kernel described in Section~\ref{sec:graph_kernel}, a kernel function $k_s:\mathcal{S}^2\to \mathbb{R}$ is defined to compute the similarity between two sequences $S_1,S_2\in \mathcal{S}$ as $k_s(S_1,S_2)$. Given a training dataset $\{(S_i, y_i)\}_{i=1}^N$, the training process solves the same optimization problem in Equation~(\ref{kernel_svm}) with $K_{ij}=k_s(S_i, S_j)$. With the learned parameters $\{\alpha_i\}_{i=1}^N$, the prediction result of an sequence $S$ is given by $\sum_{i=1}^N\alpha_i y_i k_s(S_i,S)$ and $\text{sign}\left(\sum_{i=1}^N\alpha_i y_i k_s(S_i,S)\right)$ for regression and binary classification, respectively.

Concretely, the subsequence kernel focuses on subsequences with a pre-defined length $D$. For SMILES sequences with the vocabulary $\Sigma$, the set of all possible subsequences of length $D$ is $\Sigma^D=\{U_1,U_2,\ldots,U_{|\Sigma^{D}|}\}$. For a sequence $S$, we compute
\begin{align}
    &n(U, S) = \sum_{\bm i: U=S[\bm i]} \lambda^{i_{D}-i_1+1},\\
    &\phi(S) = \left[n(U_1, S), n(U_2, S),\cdots, n(U_{|\Sigma^{D}|}, S)\right]^T,
\end{align}
where $n(U_i, G)$, $i=1,2,\ldots,|\Sigma^{D}|$ is a weighted count of $U_i$ in $S$. Specifically, $\lambda$ with $0<\lambda<1$ is introduced as a decay factor, so that a subsequence is given a smaller weight if its characters are further away from each other in the sequence $S$. With $\phi(\cdot)$, the kernel function $k_s$ between two SMILES sequences $S_1$ and $S_2$ is given by $k_s(S_1,S_2)=\phi(S_1)^T\phi(S_2)$. In practice, it is common to normalize it by $\frac{k_{s}(S_1, S_2)}{\sqrt{k_{s}(S_1, S_1)\cdot k_{s}(S_2, S_2)}}$.

\subsection{Evaluation metrics}\label{subsec: metrics}

As the molecular property prediction problem can be formulated as a regression or a binary classification task, we use different evaluation metrics for different datasets.

\subsubsection{Regression tasks}\label{subsubsec: reg_metrics}

Given a testing dataset of $N$ samples for a regression task, we denote the ground-truth labels as $\{y_i\}_{i=1}^n$ and the corresponding prediction results as $\{\hat{y}_i\}_{i=1}^N$, where $y_i, \hat{y}_i \in \mathbb{R}$. The evaluation metric is mean absolute error (MAE)~\cite{willmott2005advantages}:
\begin{equation}\label{eqn:mae}
MAE(\{y_i\}_{i=1}^N,\{\hat{y}_i\}_{i=1}^N)=\frac{1}{N}\sum_{i=1}^N |y_i-\hat{y}_i|,
\end{equation}
or root-mean-square error (RMSE):
\begin{equation}\label{eqn:rmse}
RMSE(\{y_i\}_{i=1}^N,\{\hat{y}_i\}_{i=1}^N)=\sqrt{\frac{1}{N}\sum_{i=1}^N (y_i-\hat{y}_i)^2}.
\end{equation}
A smaller MAE or RMSE indicates a better performance.

\subsubsection{Binary classification tasks}\label{subsubsec: cls_metrics}

Given a testing dataset of $N$ samples for a binary classification task, we denote the ground-truth labels as $\{y_i\}_{i=1}^n$ and the corresponding predicted classification scores as $\{\hat{y}_i\}_{i=1}^N$, where $y_i \in \{-1,1\}$ and $\hat{y}_i \in [0,1]$. The predicted classification score $\hat{y}_i$ can be transformed into the predicted label $\tilde{y}_i \in \{-1,1\}$ with a threshold $\gamma$, where $\tilde{y}_i = 1$ if $\hat{y}_i \geq \gamma$ and $\tilde{y}_i = -1$ if $\hat{y}_i < \gamma$.

If $\tilde{y}_i=y_i$ and $\tilde{y}_i=1$ (or $\tilde{y}_i=-1$), we say the $i$-th prediction is a true positive (or true negative); if $\tilde{y}_i\ne y_i$ and $\tilde{y}_i=1$ (or $\tilde{y}_i=-1$), we say the $i$-th prediction is a false positive (or false negative). With the threshold $\gamma$, we use $TP_\gamma, TN_\gamma, FP_\gamma, FN_\gamma$ to represent the number of true positives, true negatives, false positives and false negatives, respectively. And we can compute the true positive rate ($TPR$), false positive rate ($FPR$), and positive predictive value ($PPV$) under $\gamma$ as:
\begin{align}
    &TPR_\gamma=\frac{TP_\gamma}{TP_\gamma+FN_\gamma},\\
    &FPR_\gamma=\frac{FP_\gamma}{FP_\gamma+TN_\gamma},\\
    &PPV_\gamma=\frac{TP_\gamma}{TP_\gamma+FP_\gamma}.
\end{align}
Note that $TPR$ and $PPV$ are also known as recall and precision.

The evaluation metric for binary classification tasks is the area under curves (AUC) of the receiver operating characteristic (ROC) curves~\cite{fawcett2006introduction} or the precision-recall curves (PRC)~\cite{davis2006relationship}, denoted by ROC-AUC and PRC-AUC. Concretely, by varying $\gamma \in [0,1]$, the ROC curve and PRC plot how $(FPR, TPR)$ and (recall, precision) change in a $xy$-coordinate plane, respectively. A larger ROC-AUC or PRC-AUC indicates a better performance.

\subsection{Training losses}\label{sec:loss}

We introduce the loss functions to train the deep learning components of AdvProp, \emph{i.e.}, our proposed ML-MPNN and contrastive-BERT. Note that the contrastive-BERT has an extra one-time self-supervised pre-training phase with a contrastive loss, as described in Section~\ref{sec:contrastive-Bert}. The fine-tuning phase of the contrastive-BERT uses the same supervised losses as ML-MPNN. For all the training of deep learning models, we employ the batch training, where we randomly select a batch of $B$ samples from the training dataset in each training iteration. The Adam\cite{kingma:adam} optimizer is used to minimize the loss.

\subsubsection{Supervised loss for regression tasks}

In regression tasks, we use the mean squared error (MSE) loss:
\begin{align}
    L_{MSE}(\{y_i\}_{i=1}^B,\{\hat{y}_i\}_{i=1}^B)=\frac{1}{B}\sum_{i=1}^B(y_i-\hat{y}_i)^2,
\end{align}
where $y_i,\hat{y}_i\in \mathbb{R}$ are the ground-truth label and prediction result of the $i$-th training sample in the batch, respectively.

\subsubsection{Binary cross entropy loss}

For all binary classification tasks other than the AI Cures challenge, we only use the binary cross entropy (BCE) loss:
\begin{align}
    L_{BCE}(\{y_i\}_{i=1}^B,\{\hat{y}_i\}_{i=1}^B)=-\frac{1}{B}\sum_{i=1}^B\left(\mathbbm{1}_{\{y=1\}}(y_i)\log(\hat{y}_i)+ \mathbbm{1}_{\{y=-1\}}(y_i)\log(1-\hat{y}_i)\right),
\end{align}
where $y_i\in\{-1,1\}$ is the ground-truth label of the $i$-th training sample in the batch, $\hat{y}_i\in(0,1)$ is the corresponding predicted classification score, and $\mathbbm{1}_{\{y=1\}}(\cdot)$ and $\mathbbm{1}_{\{y=-1\}}(\cdot)$ are indicator functions.

\subsubsection{Stochastic optimization of AUC losses}

For the AI Cures challenge, in addition to the BCE loss, we also employ two methods for maximizing ROC-AUC and PRC-AUC, respectively. We briefly describe two methods below and refer the audience to the original papers~\cite{yuan2020robust, qi2021stochastic} for more details. 

To maximize the ROC-AUC, we use the AUC margin loss~\cite{yuan2020robust}. The problem is formulated as a min-max optimization problem:
\begin{align}
	\min_{w,a,b}\max_{\alpha\ge 0} L_{AUC-M}(\{y_i\}_{i=1}^B,\{\hat{y}_i\}_{i=1}^B)
	=\frac{1}{B}\sum_{i=1}^B&\left\{(1-p)\mathbbm{1}_{\{y=1\}}(y_i)(\hat{y}_i-a)^2+p\mathbbm{1}_{\{y=-1\}}(y_i)(\hat{y}_i-b)^2-p(1-p)\alpha^2\right. \nonumber \\
	&+\left.2\alpha\left[p(1-p)m_1+p\mathbbm{1}_{\{y=-1\}}(y_i)\hat{y}_i-(1-p)\mathbbm{1}_{\{y=1\}}(y_i)\hat{y}_i\right]\right\},
\end{align}

 where $\hat{y}_i$ is the prediction score output by the neural network parameterized by $w$, $p$ is the prior probability that the data belongs to the positive class, $m_1>0$ is a margin hyper-parameter. For solving the min-max problem, we employ the PESG method based on mini-batching~\cite{yuan2020robust}. 
 
 To maximize the PRC-AUC loss, we approximate it by average precision (AP) and use the following surrogate loss to compute AP (for maximization):
 \begin{align}
 	L_{AP}(\{y_i\}_{i=1}^B,\{\hat{y}_i\}_{i=1}^B)=\frac{1}{B_+}\sum_{i=1}^B\frac{\sum_{j=1}^B\mathbbm{1}_{\{y=1\}}(y_j)\left(\max\{m_2-\hat{y}_i+\hat{y}_j, 0\}\right)^2}{\sum_{j=1}^B\left(\max\{m_2-\hat{y}_i+\hat{y}_j, 0\}\right)^2}\mathbbm{1}_{\{y=1\}}(y_i),
 \end{align}
 where $\hat{y}_i$ is the prediction score output by the neural network parameterized by $w$, $B_+$ is the number of positive examples, $m_2$ is a tunable hyperparameter. For solving the above problem, we employ the SOAP-ADAM method~\cite{qi2021stochastic}. 
 
 For implementing both PESG optimizing the AUC margin loss and SOAP-ADAM for optimizing the AP surrogate loss, we use the code in LibAUC package~\cite{libauc}. 

\subsubsection{Self-supervised loss for the masked embedding recovery task}
The masked embedding recovery task uses a self-supervised contrastive loss. Given a SMILES sequence $S$, the input and output embeddings of the contrastive-BERT are $h_i$ and $o_i$, $i=1,2,\ldots,|S|$, respectively. If $S[i]$ is masked, the corresponding contrastive loss is given by
\begin{equation}\label{eqn:pre_train_loss}
    L_{contrast}(o_i, h_1, h_2, \ldots, h_{|S|}) = -\log \frac{\exp(sim(o_i,h_i)/\tau)} {\sum_{j=1}^{|S|} \exp(sim(o_i,h_j)/\tau)},
\end{equation}
where $\tau$ is a temperature parameter and the similarity function $sim(\cdot, \cdot)$ computes the cosine similarity, defined by
\begin{equation}
sim(o, h) = \frac{o^T h}{|o| \cdot |h|}.
\end{equation}
If a SMILES sequence has multiple masked characters, the final loss is the averaged contrastive loss.

\subsection{Configurations}\label{subsec: config}


The settings of our device are - GPU: Nvidia GeForce RTX 2080 Ti 11GB; CPU: Intel Xeon Silver 4116 2.10GHz; OS: Ubuntu 16.04.3 LTS.

We implement our proposed ML-MPNN in PyTorch~\cite{paszke2019pytorch} and Pytorch Geometric~\cite{Fey/Lenssen/2019}. Following D-MPNN~\cite{yang2019analyzing}, we incorporate additional global molecular features that can be easily obtained with the open-source package RDKit~\cite{rdkit}. Each molecule has a 200-dimensional global molecular feature, which is concatenated with the final graph-level representation learned by our ML-MPNN as inputs to a 2-layer MLP to make the prediction. The task-specific configurations include the settings of initial learning rate, learning rate decay factor, weight decay, hidden dimension, dropout rate, and batch size, as listed in Supplementary Table~7.

Our contrastive-BERT model is also implemented in PyTorch. The network is composed of 6 Transformer layers with 1024 hidden units and 4 attention heads. For the one-time pre-training phase, we train the model for 10 epochs, with the initial learning rate set to 0.0001 and decay with the cosine annealing strategy. The batch size is set to 128 and the dropout rate is 0.1. In the fine-tuning phase, the settings are task-specific. Specifically, we tune batch size, the number of training epochs and learning rate for each dataset, as listed in Supplementary Table~8.

As for kernel methods, the number of iterations of graph kernels, the subsequence length and the decay factor of string kernels, and whether kernel normalization is used are task-specific. The setting of these hyperparameters for each dataset is summarized in Supplementary Table~9.

\subsection*{Reporting summary}

Further information on research design can be found in the Nature Research Reporting Summary linked to this article.

\section*{Data availability}


The dataset for AI Cures open challenge task~\cite{aicures} is available at \url{https://www.aicures.mit.edu/data}. The dataset for antibiotic discovery, provided by Stokes et al.~\cite{stokes2020deep}, can be found at \url{https://www.sciencedirect.com/science/article/pii/S0092867420301021}. The datasets and splits of MoleculeNets benchmarks~\cite{wu2018moleculenet} for molecular property prediction can be downloaded from \url{http://deepchem.io/datasets/MolNet_bkp.zip} and \url{http://deepchem.io/trained_models/Hyperparameter_MoleculeNetv3.tar.gz}. The dataset of $2$ million molecules from ZINC~\cite{irwin2012zinc}, prepared by Hu et al.~\cite{hu2019strategies}, for pretraining our contrastive-Bert can be downloaded from \url{http://snap.stanford.edu/gnn-pretrain/data/chem_dataset.zip}.

We also collect datasets and corresponding splits of MoleculeNet benchmarks and the dataset of $2$ million molecules from ZINC at \url{https://github.com/divelab/MoleculeX/tree/master/AdvProp/datasets} in our AdvProp tool. We recommend our users to use this collection for convenience.

\section*{Code availability}


The code for AdvProp training, prediction and evaluation (in Python/PyTorch), is publicly available at \url{https://github.com/divelab/MoleculeX}. We use the LibAUC library~\cite{libauc}
for stochastic optimization of deep learning models with advanced AUC loss functions, including ROC-AUC loss and PRC-AUC loss.


\end{document}